\documentclass[prb,twocolumn,showpacs,floatfix,superscriptaddress]{revtex4}
\usepackage{graphicx}
\usepackage{amsmath}
\usepackage{amsfonts}

\begin{document}

\title{Current reversal in collective ratchets induced by lattice instability}
\author{L. Dinis}
\affiliation{Departamento\ de F{\'{\i}}sica At{\'{o}}mica, Molecular y Nuclear and GISC,
Universidad Complutense de Madrid, 28040-Madrid, Spain}
\author{E.M. Gonz\'{a}lez}
\affiliation{Departamento\ de F{\'{\i}}sica de Materiales, Universidad Complutense de
Madrid, 28040-Madrid, Spain}
\author{J.V. Anguita}
\affiliation{Instituto de Microelectr\'{o}nica de Madrid, Consejo Superior de
Investigaciones Cient\'{\i}ficas, Tres Cantos, 28670-Madrid, Spain}
\author{J.M.R. Parrondo}
\affiliation{Departamento\ de F{\'{\i}}sica At{\'{o}}mica, Molecular y Nuclear and GISC,
Universidad Complutense de Madrid, 28040-Madrid, Spain}
\author{J.L. Vicent}
\affiliation{Departamento\ de F{\'{\i}}sica de Materiales, Universidad Complutense de
Madrid, 28040-Madrid, Spain}
\date{\today }

\begin{abstract}
A collective mechanism for current reversal in superconducting vortex
ratchets is proposed. The mechanism is based on a two-dimensional
instability of the ground state ($T=0$) of the system. We illustrate our results
with numerical simulations and experiments in Nb superconducting films
fabricated on top of Si substrates with artificially induced asymmetric
pinning centers.
\end{abstract}

\maketitle
\section{Introduction}
Rectification of motion and fluctuations  in the nanoscale is
becoming a major field of research \cite{peter}. Rectifying
mechanisms or \emph{ratchets} have been used to explain how protein
motors
work \cite{rnap,kinesin} and to design new separation techniques \cite{separation} or synthetical chemical motors \cite{leigh,aaron}.
Superconductors have become a powerful tool to study ratchet mechanisms 
\cite{zapata}. Recently, a superconducting vortex ratchet device has been
reported by Villegas et al \cite{science}. In that experiment, the
rocking ratchet mechanism is due to a superconducting film patterned
with a lattice of asymmetric potentials acting as pinning centers.
An input ac current yields an output dc voltage in the
superconducting film, revealing a rich phenomenology of single and
multiple current reversals \cite{science,Souza,dinis}, very
sensitive to the underlying vortex dynamics. The interest of vortex
ratchets is then twofold: it reveals new collective rectification
mechanisms and
 sheds light on the physics of vortices in superconductors.

Current reversals in vortex ratchets have been explained, in the
framework of one-dimensional models, by the coexistence of pinned
and interstitial vortices moving in opposite directions
\cite{science} or by the interaction
between vortices within the pinning centers in one-dimensional channels 
\cite{Souza}. Olson and Reichhardt \cite{Olson} have studied numerically
a two-dimensional model, obtaining current reversal only when
interstitial vortices are present in the ground state. They provide
intuitive explanations of different rectification mechanisms based
on local interactions between vortices and pinning centers.

In this Letter we experimentally show that current reversal can also
occur when interstitial vortices are absent in the ground state (T=0).
Remarkably, current reversal disappears increasing either the
pinning strength or the temperature. We also present numerical
simulations of vortices as interacting Brownian particles in two
dimensions, indicating that this current reversal is due to a new
collective effect: an instability of the ground state, selective to
the sign of the applied force. The influence of lattice
instabilities on rectification has been also recently analyzed by Lu
et al. \cite{lu}, for two dimensional vortices in a substrate with a
one-dimensional modulation.

\section{Experimental results}

For our experiments, two types of asymmetric pinning centers have
been fabricated: magnetic (Ni) nanotriangles and non-magnetic (Cu)
nanotriangles. The vortex pinning force is enhanced by magnetic
centers in comparison with non-magnetic centers \cite{magnomag}. The
nanotriangles were fabricated using e-beam lithography techniques
and Si (100) wafers as substrates. The Ni or Cu arrays of
nanotriangles, on top of the substrate, are covered with a sputtered
Nb thin film. Ni or Cu thickness (triangles height) is 40 nm and Nb
film is 100 nm thick. Further details on this fabrication technique
can be found elsewhere \cite{Mart:98}. For the present work, we have
fabricated arrays with the same nanotriangle dimensions and array
periodicity than those in Ref.~\onlinecite{science}.

We have measured magnetotransport in these films using a commercial He
cryostat. The variable temperature insert allows controlling temperature
with stability of 1 mK. For these experiments, samples were patterned with a
cross-shaped measuring bridge \cite{science}, by using optical lithography
and ion-etching. This patterned bridge allows us to control the Lorentz
force on the vortices in the mixed state: taking into account 
$\vec{F}_{L}=\vec{J}\times \vec{n}\phi _{0}$ 
(with $\phi _{0}=2.07\times 10^{-15}$ Wb and $\vec{n}$ a unitary vector 
parallel to the applied magnetic field). On the
other hand, from the expression for the electric field $\vec{E}=\vec{B}%
\times \vec{v}$, where $\vec{B}$ is the applied magnetic field and $\vec{v}$
the vortex-lattice velocity, we can calculate this velocity 
$v=V/\left( {dB}\right) $ from the measured voltage drops $V$ ($d$
being the distance between contacts). See Ref.~\onlinecite{science} for more
experimental details.

The dc magnetoresistance in the mixed state of samples with periodic arrays
of pinning centers exhibits well-known commensurability phenomena 
\cite{Daldini:74,martin:97}, in which minima develop as a consequence of
geometrical matching between the vortex-lattice and the underlying
periodic structure. These minima are equally spaced (two neighbor
minima are always separated by the same magnetic field value). For
example, in the case of square arrays of nanostructured pinning
centers, minima appear at applied magnetic fields
$H_{\mathrm{m}}=n(\phi _{0}/a^{2})$, where $a$ is the lattice
parameter of the square array. Hence, the number of vortices $n$ per
array unit cell can be known by simple inspection of the dc
magnetoresistance $R(H)$ curves. Moreover, for non-magnetic pinning
centers Mkrtchyan and Shmidt \cite{Smidt72} have given a rough
estimation of the maximum number of vortices that can be pinned in
each center, which could confirm the matching field minima values.
This filling factor can be calculated as the ratio between half the
dimension of the pinning center (the triangle side is around 650 nm)
\cite{science} and two times the superconducting coherence length
(around 60 nm for these samples and temperatures close to the
critical temperature) \cite{martin:97}. In our samples this rough
estimation gives us approximately three vortices per triangle, in
agreement with the matching field data (see also \onlinecite{science}).
Therefore, we know, for selected values of the applied magnetic
field, how many vortices there are per unit cell and where they are,
this is, if they are interstitial vortices or vortices in the
pinning centers.

We want to underline that collective behavior of vortices in films
with periodic pinning is crucial to understand vortex lattice
reconfiguration effects \cite{Mart:1999} or vortex channeling
\cite{Velez1PRB02}. Also, collective effects in ratchets have been
shown to yield new interesting phenomena \cite{colectivo}.

\begin{figure}[tbp]
\begin{center}
\includegraphics[width=0.45\textwidth]{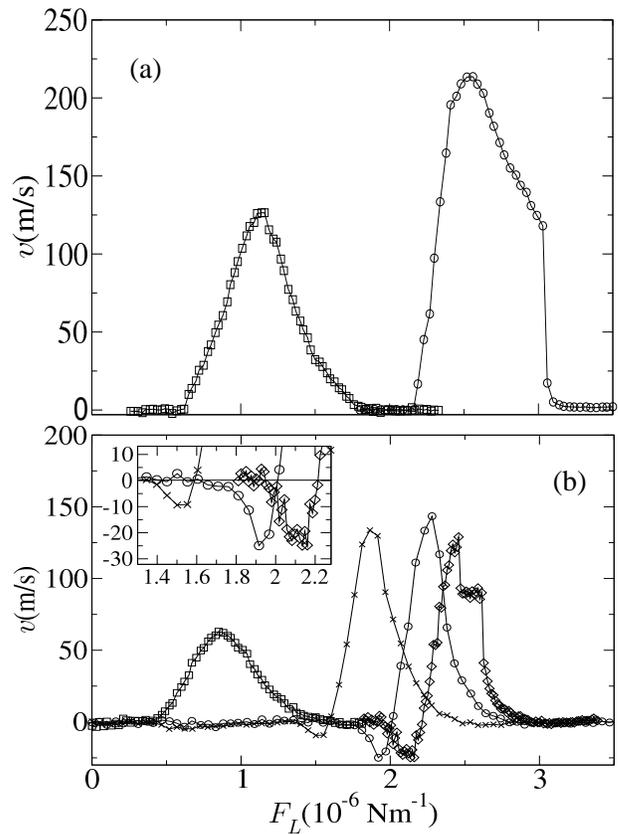}
\end{center}
\caption{Net velocity of vortices versus ac Lorentz force amplitude
($\protect\omega =10$kHz). Array periodicity 770 nm, triangle base 620 nm.
(a) Sample (A): Nb film with Ni triangles $T_{c}=8.35K$, 
$T/T_c=0.99$ ($T=8.265$K) ($\square$), $T/T_c=0.98$ ($T=8.174$K) ($\circ$). (b) Sample (B): Nb film with Cu triangles $T_{c}=8.24$K. $T=8.165$K ($\square$), $T=8.102$K ($\times $), $T=8.08$K ($\circ$), $T= 8.065$K ($\Diamond $). The negative velocity part of each curve is shown in the inset.}
\label{fig:NiCu}
\end{figure}

Measurements of the vortex lattice average velocity as a function of
the applied force are depicted in Fig.~\ref{fig:NiCu}(a) for sample
(A), Ni triangles, and three vortex per triangle ($n=3$). There is
no current reversal
in the rectified signal in agreement with the explanation given in 
Ref.~\onlinecite{science} (see below). Experiments for sample (B), Cu triangles, also for $n=3$ (Fig.~\ref{fig:NiCu}(b)) show a similar behavior for
$T=8.165$K ( $T=0.99\,T_{c}$), but a current reversal appears decreasing the temperature, despite the fact that there are still three vortices per triangle.

In the one-dimensional approach reported in Ref.~\onlinecite{science}, the
vortex lattice does not play any role, and the current reversal was
explained assuming that pinned and interstitial vortices are
rectified in opposite directions. However, this assumption fails to
explain current reversal for the matching field $n=3$, since in this
case there are no interstitial vortices in the ground state of the
system. The simulations performed by Olson and Reichhardt
\cite{Olson}  do not exhibit any current reversal for $n=3$ and the
rectification mechanisms that they propose do not apply for this
case either. Therefore, a new explanation of current reversal is
required. We have found by numerical simulations such an explanation
based on the interplay between the vortex lattice and the geometry
of the triangular defects.

\section{Theoretical model}
\subsection{Simulations}
The simulations have been performed by numerically solving Langevin
equations for the movement of the vortices
\begin{equation}
\eta \dot{\mathbf{x}}=-\partial _{x}U_{vv}+\mathbf{\nabla }V_{p}+\mathbf{F}_{\mathrm{ext}}+\xi (t).  \label{eq:langevin}
\end{equation} $\mathbf{F}_{\rm ext}$ is the Lorentz force resulting from the
applied
current, \ $U_{vv}$\ the usual vortex-vortex interaction (see 
Ref.~\cite{Tinkham}),$\ V_{p}\ $the pinning potential, $\eta$ the friction
coefficient, and $\xi (t)$ a white gaussian thermal noise. The
pinning force and vortex-vortex interaction must be in agreement
with the experimental situation of three vortices per pinning site
(triangle). The interaction of the vortices with the pinning defects
is modelled by a potential $V_{p}$ in the shape of a triangle and
with a hyperbolic tangent profile. The depth of the potential is
$V_{p0}=0.002\text{ pN}\mu \text{m}$, so that depinning forces have
typical values of the order of $10^{-6}\text{N/m}$. Experiments
\cite{VillePRB05,VanPRL05} show that the system is adiabatic in the
region of frequencies used. This allows us to obtain the expected ac
signal from the velocity-force response curves obtained in
simulations, both for constant positive and negative applied force.

\begin{figure}[tbp]
\begin{center}
\includegraphics[width=0.45\textwidth]{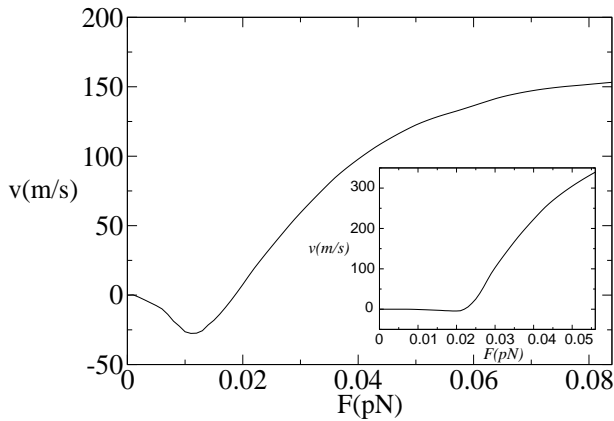}
\end{center}
\caption{Simulation results. Three vortices per triangle, pinning $0.002$ pN
$\protect\mu$m. Inset: pinning $0.0036$ pN$\protect\mu$m. }
\label{fig:n3inversion}
\end{figure}

Results are shown in Fig.~\ref{fig:n3inversion} where a window of downward
rectification can be observed. Let us stress at this point that the
\textquotedblleft natural\textquotedblright\ direction of rectification for
pinned vortices is upwards. The reason is that the vortex feels a smaller
force at the tip of the triangle than at the triangle base. The force at the
tip can be, for an equilateral triangle, as lower as half the force at the
base of the triangle.

\subsection{Ground state instability}
\begin{figure}[tbp]
\begin{center}
\includegraphics[width=.42\textwidth]{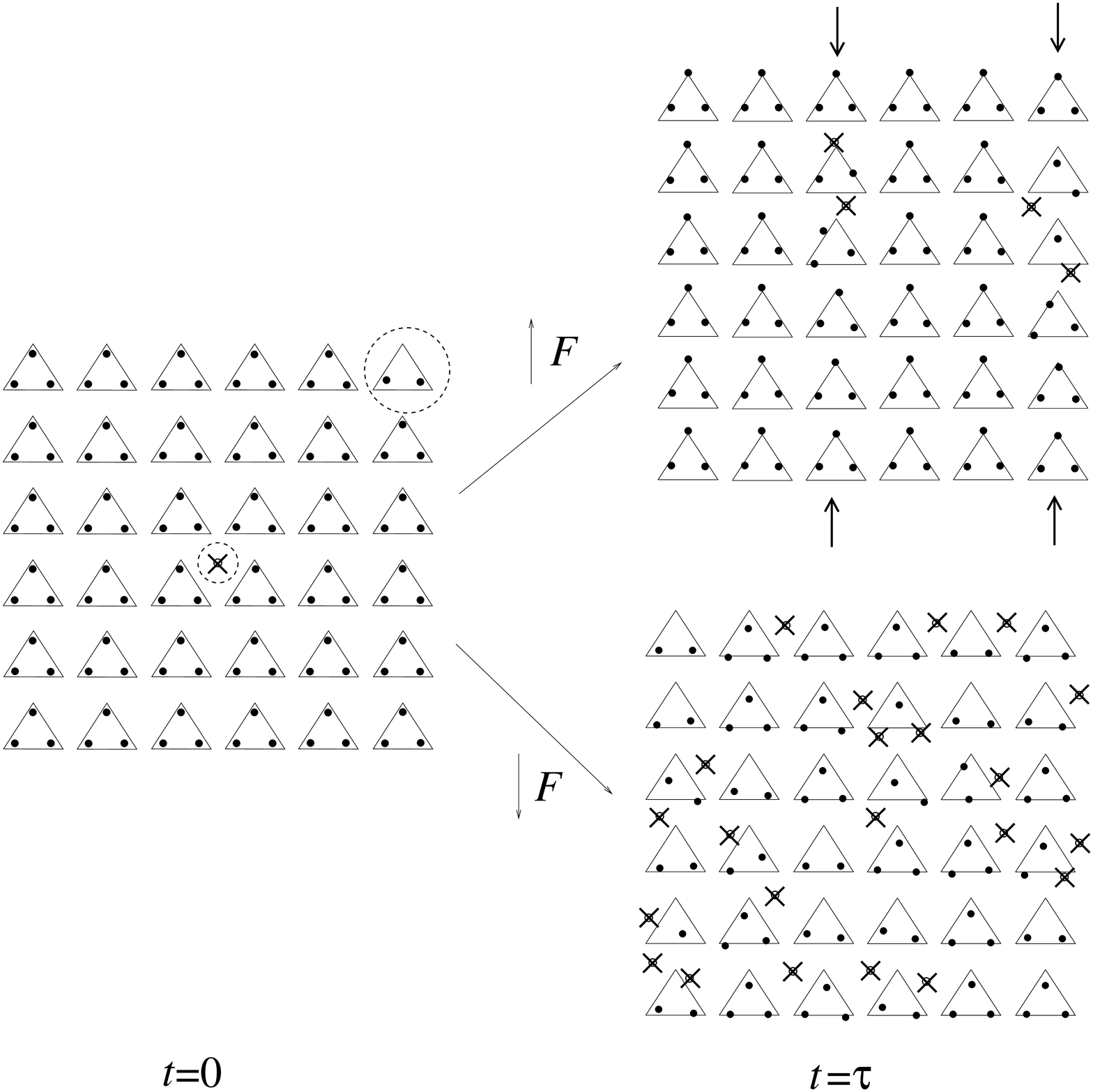}
\end{center}
\caption{Downward rectification mechanism at $T=0$ K: snapshots from
simulations. Initial condition (left) and configurations after evolution 
($\protect\tau=6.5\times 10^{-9}$s) with positive force $F=0.032$pN (right,
up) and negative force $-F$ (right, down). Dashed circles show the defective
triangle and the only interstitial present in the initial condition. Thick
arrows point out the only two columns presenting any motion when positive
force is applied. Comparison of the right panels shows that the number of interstitial vortices ($\times$) dramatically increases when the external force points downwards.}
\label{fig:snapshots}
\end{figure}

What is then the origin of the downward rectification for small forces? Our
simulations indicate that the rectification is due to a two-dimensional
instability of the ground state under small downward forces. The ground
state of the system consists of vortices located in the corners of each
triangle, without interstitial vortices. However, for finite temperature,
there are some \textquotedblleft defective triangles\textquotedblright\ in
the actual configuration of vortices, i.e., some triangles which have only
two vortices inside. Accordingly, some interstitial vortices will be
randomly spread along the sample. Fig.~\ref{fig:snapshots} shows one of
these configuration with only one defective triangle out of $6\times 6$. For
low forces, motion is induced by the interstitial vortices \emph{both} in
the downward and upward direction. However, contrary to the picture
presented in previous works \cite{science}, there is not a continuous motion
of interstitial vortices along the space between triangles: the interstitial
vortex enters the nearest triangle expelling one of the three vortices
inside.

The aforementioned instability, selective to the sign of the
external force, is shown in Fig.~\ref{fig:snapshots}. We have chosen
an initial condition with only one interstitial vortex and one
defective triangle, located at some distance. In the right figures
we have plotted the configurations of vortices after an upward and
downward force has been applied for $\tau =6.5\times 10^{-9}$
seconds (long enough for a depinned vortex to cover the whole sample
several times), respectively. This simulations evolved at zero
temperature to show the mechanism in a clearer manner. We see that,
in the case of the upward force, the interstitial vortex remains in
\emph{one single column}. The column with a defective triangle also
moves, but again vortices remain in that column. As a consequence,
there is a positive current of vortices but the motion is
constrained to two columns. When the interstitial vortex enters a
triangle, the top vortex in the triangle moves upwards, out of it
and into the following one. In addition to that, in the defective
triangle, one of the two vortices always escapes through the tip,
becoming and interstitial vortex and triggering a process similar to
that in the column with an extra vortex. Both motions propagate
along the column without disturbing the neighboring columns.

On the other hand, under a downward force, the initial defect in the
ground state propagates along the whole sample, as it is clearly
shown in Fig.~\ref{fig:snapshots}. A more detailed analysis of the
simulations show that, when an interstitial vortex enters a
triangle, one of the two bottom vortices is expelled but now can
move to a triangle \emph{in one of the nearest columns}. It even
happens frequently that the two vortices in the base of the triangle
are depinned. Consequently, the initial defect is then spread out
along the horizontal direction, yielding a considerable large
fraction of depinned vortices which increase the overall motion in
the system, yielding a net downward rectification and the
corresponding  current reversal.

As was noted before, current inversion may disappear when the
pinning potential is increased. This effect was observed in
experiments where the Cu triangles were replaced by Ni ones which
have a higher pinning strength (see Fig.~\ref{fig:NiCu}(a)), and it
is also reproduced in our simulations. Results for pinning potential
$V_{p0}=0.0036\text{pN}\mu \text{m}$ do not show current reversal,
as depicted in the inset in Fig.~\ref{fig:n3inversion}. The
increased pinning strength implies that a larger force is needed to
depin vortices when pushing downwards. According to simulations,
when the same force is applied upwards, the aforementioned columnar
movement starts, but also neighboring columns without defective
triangles or extra vortices depin, and eventually motion spreads all
over the sample. In other words, for these moderate values of the
force, the instability appears in both directions.

\begin{figure}[tbp]
\begin{center}
\includegraphics[width=0.43\textwidth]{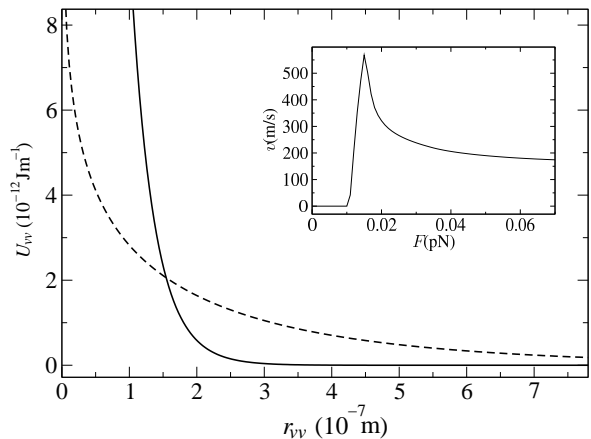}
\end{center}
\caption{Vortex-vortex interaction versus vortex-vortex distance. Solid
line: interaction used in simulations corresponding to current reversal (see
Fig.~\protect\ref{fig:n3inversion}, pinning 
$V_{p0}=0.002$pN$\protect\mu$m.). Dashed line: interaction used in simulations showing no current
reversal, which are depicted in the inset (pinning $V_{p0}=0.002$pN
$\protect\mu$m). }
\label{fig:k0inset}
\end{figure}

Finally, current reversal in sample (B), Cu triangles, also vanishes
when temperature is raised (see Fig.~\ref{fig:NiCu}(b)). We
can explain this behavior taking into account that the penetration
depth of the superconductor increases with temperature. As a result,
the vortex-vortex interaction strength, $U_{vv}$, decreases at short
distance but its range becomes longer \cite{Tinkham}, as depicted in
Fig.~\ref{fig:k0inset}. In this case we observe that the long range
vortex lattice order precludes the instability responsible for
current reversal (see the inset in Fig.~\ref{fig:k0inset}).

\section{Conclusions}
In summary, our simulations show that the asymmetric substrate
induces an instability sensitive to the direction of the external
force, affecting rectification. This effect can explain the current
reversal observed in our experiments for $n=3$. Moreover, our work
indicates that the lattice configuration, and consequently its
dynamical properties, can be controlled by external forces. This
interplay between rectification, driving forces and lattice
configuration can induce other interesting phenomena such as
transitions between different lattice configurations \cite{dinis},
and could help to design new rectifying devices, not only in
superconducting films, but also in other two dimensional collective
systems, such as Josephson arrays or colloidal suspensions.

We acknowledge support by Spanish Ministerio de Educaci\'on y
Ciencia (NAN04-09087, FIS05-07392, MAT05-23924E, MOSAICO), CAM
(S-0505/ESP/0337) and UCM-Santander. Computer simulations were
performed at ``Cluster de c\'alculo para T\'ecnicas F\'isicas'' of
UCM, funded in part by UE-FEDER program and in part by UCM and in
the ``Aula Sun Microsystems'' at UCM.

\end{document}